\documentclass[citeautoscript,floatfix,superscriptaddress,twocolumn,showpacs,prb,aps,10pt]{revtex4-1}

\usepackage{graphicx} \usepackage{amsmath, amssymb, bm}
\usepackage{dcolumn} \usepackage{color}
\usepackage{eso-pic}
\usepackage{fix-cm}
\usepackage[english]{babel}

\usepackage{multirow}
\usepackage{sidecap}

\usepackage{marginnote}
\usepackage{ulem} 

\newcommand{\sss}{\scriptscriptstyle}
\newcommand{\sst}{\scriptstyle}

\newcommand{\stext}[1]{\sss \text{#1} \sst}

\renewcommand{\emph}[1]{\textit{#1}}

\begin{document}
\title{Anisotropy, phonon modes, and lattice anharmonicity from dielectric function tensor analysis of monoclinic cadmium tungstate}

\author{A. Mock}
\email{amock@huskers.unl.edu}
\homepage{http://ellipsometry.unl.edu}
\affiliation{Department of Electrical and Computer Engineering and Center for Nanohybrid Functional Materials, University of Nebraska-Lincoln, U.S.A.}
\author{R.~Korlacki}
\affiliation{Department of Electrical and Computer Engineering and Center for Nanohybrid Functional Materials, University of Nebraska-Lincoln, U.S.A.}
\author{S.~Knight}
\affiliation{Department of Electrical and Computer Engineering and Center for Nanohybrid Functional Materials, University of Nebraska-Lincoln, U.S.A.}
\author{M. Schubert}
\affiliation{Department of Electrical and Computer Engineering and Center for Nanohybrid Functional Materials, University of Nebraska-Lincoln, U.S.A.}
\affiliation{Leibniz Institute for Polymer Research, Dresden, Germany}
\affiliation{Department of Physics, Chemistry, and Biology (IFM), Link{\"o}ping University, SE 58183, Link{\"o}ping, Sweden}

\date{}

\begin{abstract}
We determine the frequency dependence of four independent CdWO$_4$ Cartesian dielectric function tensor elements by generalized spectroscopic ellipsometry within mid-infrared and far-infrared spectral regions. Single crystal surfaces cut under different angles from a bulk crystal, (010) and (001), are investigated. From the spectral dependencies of the dielectric function tensor and its inverse we determine all long wavelength active transverse and longitudinal optic phonon modes with $A_u$ and $B_u$ symmetry as well as their eigenvectors within the monoclinic lattice. We thereby demonstrate that such information can be obtained  completely without physical model line shape analysis in materials with monoclinic symmetry. We then augment the effect of lattice anharmonicity onto our recently described dielectric function tensor model approach for materials with monoclinic and triclinic crystal symmetries [Phys. Rev. \textbf{B}, 125209 (2016)], and we obtain excellent match between all measured and modeled dielectric function tensor elements. All phonon mode frequency and broadening parameters are determined in our model approach. We also perform density functional theory phonon mode calculations, and we compare our results obtained from theory, from direct dielectric function tensor analysis, and from model lineshape analysis, and we find excellent agreement between all approaches. We also discuss and present static and above reststrahlen spectral range dielectric constants. Our data for CdWO$_4$ are in excellent agreement with a recently proposed generalization of the Lyddane-Sachs-Teller relation for materials with low crystal symmetry [Phys. Rev. Lett. \textbf{117}, 215502 (2016)].
\end{abstract}
 \maketitle

\section {Introduction}

Metal tungstate semiconductor materials (AWO$_4$) have been extensively studied due to their remarkable optical and luminescent properties. Because of their properties, metal tungstates are potential candidates for use in phosphors, in scintillating detectors, and in optoelectronic devices including lasers.\cite{Mikhailik_2005,Blasse_2012,Kato_2005} Tungstates usually crystallize in either the tetragonal scheelite or monoclinic wolframite crystal structure for large (A = Ba, Ca, Eu, Pb, Sr) or small (A = Co, Cd, Fe, Mg, Ni, Zn) cations, respectively.\cite{Lacomba-Perales_2008}  The highly anisotropic monoclinic cadmium tungstate (CdWO$_4$) is of particular interest for scintillator applications, because it is non-hygroscopic, has high density (7.99~g/cm$^3$) and therefore high X-ray stopping power\cite{Blasse_2012}, its emission centered near 480~nn falls within the sensitive region of typical silicon-based CCD detectors\cite{Banhart_2008}, and its scintillation has high light yield (14,000~photons/MeV) with little afterglow\cite{Blasse_2012}. Raman spectra of CdWO$_4$ have been studied extensively\cite{Blasse_1975,Daturi_1997,Gabrusenoks_2001,Lacomba-Perales_2009,Ruiz-Fuertes_2011}, and despite its use in detector technologies, investigation into its fundamental physical properties such as optical phonon modes, and static and high-frequency dielectric constants is far less exhaustive.

\begin{figure}[!hbt]
  \begin{center}
  \includegraphics[width=0.99\linewidth,natwidth=2138,natheight=1300]{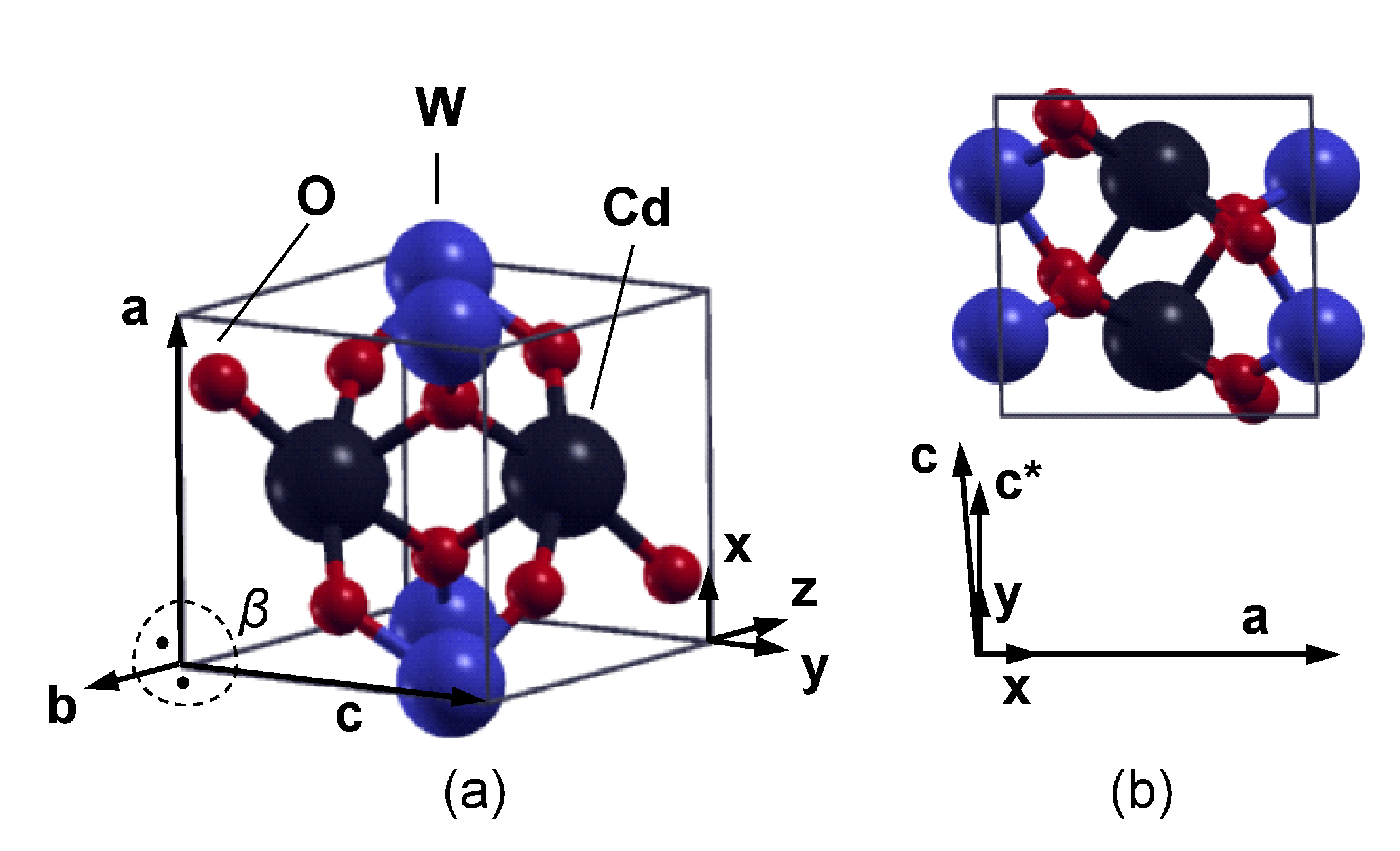}
    \caption{(a) Unit cell of CdWO$_4$, monoclinic angle $\beta$, and Cartesian coordinate system ($x$, $y$, $z$) used in this work. (b) View onto the $\mathbf{a}$ - $\mathbf{c}$ plane along axis $\mathbf{b}$, which points into the plane. Indicated is the vector $\mathbf{c^{\star}}$, defined for convenience here. See section \ref{dtmsection} for details.}
    \label{fig:CdWO4cell}
  \end{center}
\end{figure}

Infrared (IR) spectra was reported by Nyquist and Kagel, however, no analysis or symmetry assignment was included.\cite{Nyquist_1971} Blasse\cite{Blasse_1975} investigated IR spectra of HgMoO$_4$ and HgWO$_4$ and also reported analysis of CdWO$_4$ in the spectral range of 200-900~cm$^{-1}$ and identified 11 IR active modes but without symmetry assignment. Daturi \textit{et al.}\cite{Daturi_1997} performed Fourier transform IR measurements of
CdWO$_4$ powder. An incomplete set of IR active modes was identified, and a tentative symmetry assignment was provided.  A broad feature between 260-310~cm$^{-1}$ remained unexplained. Gabrusenoks \textit{et al.}\cite{Gabrusenoks_2001} utilized unpolarized far-IR (FIR) reflection measurements from 50-5000~cm$^{-1}$ and identified 7 modes with $B_u$ symmetry but did not provide their frequencies. Jia \textit{et al.}\cite{Jia_2007} studied CdWO$_4$ nanoparticles using
FT-IR between 400-1400~cm$^{-1}$, and identified 6 absorption peaks in this range without symmetry assignment. Burshtein~\textit{et al.}\cite{Burshtein} utilized infrared reflection spectra and identified 14 IR active modes along with symmetry assignment. Lacomba-Perales \textit{et al.}\cite{Lacomba-Perales_2009} studied phase transitions in CdWO$_4$ at high pressure
and provided results of density functional theory calculations for all long-wavelength active modes. Shevchuk and Kayun\cite{Shevchuk_2007} reported on the effects of temperature on the dielectric permittivity of single crystalline (010) CdWO$_4$ at 1~kHz yielding a value of approximately 17 at room temperature. Many of these studies were conducted on the (010) cleavage plane of CdWO$_4$, and therefore, the complete optical response due to anisotropy in the monoclinic crystal symmetry was not investigated. However, in order to accurately describe the full set of phonon modes as well as static and high-frequency dielectric constants of monoclinic CdWO$_4$, a full account for the monoclinic crystal structure must be provided, both during conductance of the experiments as well as during data analysis. Overall, up to this point, the availability of accurate phonon mode parameters and dielectric function tensor properties at long wavelengths for CdWO$_4$ appears rather incomplete.

In this work we provide a long wavelength spectroscopic investigation of the anisotropic properties of CdWO$_4$ by Generalized Spectroscopic Ellipsometry (GSE). We apply our recently developed model for complete analysis of the effects of long wavelength active phonon modes in materials with monoclinic crystal symmetry, which we have demonstrated for a similar analysis of $\beta$-Ga$_2$O$_3$.\cite{Schubert_2016} Our investigation is augmented by density functional theory calculations. Ellipsometry is an excellent non-destructive technique, which can be used to resolve the state of polarization of light reflected off or transmitted through samples, both real and imaginary parts of the complex dielectric function can be determined at optical wavelengths~\cite{Drude87,Drude88,Drude_1900}. Generalized ellipsometry extends this concept to arbitrarily anisotropic materials and, in principle, allows for determination of all 9 complex-valued elements of the dielectric function tensor~\cite{SchubertADP15_2006}.
Jellison \textit{et al.} first reported generalized ellipsometry analysis of a monoclinic crystal, CdWO$_4$, in the spectral region of 1.5 -- 5.6~~eV.\cite{JellisonPRB2011CdWO4} It was shown that 4 complex-valued dielectric tensor elements are required for each wavelength, which were determined spectroscopically, and independently of physical model line shape functions. Jellison~\textit{et al.} suggested to use 4 independent spectroscopic dielectric function tensor elements instead of the 3 diagonal elements used for materials with orthorhombic, hexagonal, tetragonal, trigonal, and cubic crystal symmetries. Recently, we have shown this approach in addition to a lineshape eigendisplacement vector approach applied to $\beta$-Ga$_2$O$_3$.\cite{Schubert_2016} We have used a physical function lineshape model first described by Born and Huang,~\cite{Born54} which uses 4 interdependent dielectric function tensor elements for monoclinic materials. The Born and Huang model permitted determination of all long wavelength active phonon modes, their displacement orientations within the monoclinic lattice, and the anisotropic static and high-frequency dielectric permittivity parameters. Here, we investigate the dielectric tensor of CdWO$_4$ in the FIR and mid-IR (MIR) spectral regions. Our goal is the determination of all FIR and MIR active phonon modes and their eigenvector orientations within the monoclinic lattice. In addition, we determine the static and high-frequency dielectric constants. We use generalized ellipsometry for measurement of the highly anisotropic dielectric tensor.  Furthermore, we observe and report in this paper the need to augment anharmonic broadening onto our recently described model for polar vibrations in materials with monoclinic and triclinic crystal symmetries.\cite{Schubert_2016} With the augmentation of anharmonic broadening we are able to achieve a near perfect match between our experimental data and our model calculated dielectric function spectra. In particular, in this work we exploit the inverse of the experimentally determined dielectric function tensor and directly obtain the frequencies of the longitudinal phonon modes. We also demonstrate the validity of a recently proposed generalization of the Lyddane-Sachs-Teller relation\cite{Lyddane41} to materials with monoclinic and triclinic crystal symmetries\cite{Schubert_2016_LST} for CdWO$_4$. We also demonstrate the usefulness of the generalization of the dielectric function for monoclinic and triclinic materials in order to directly determine frequency and broadening parameters of all long wavelength active phonon modes regardless of their displacement orientations within CdWO$_4$. This generalization as a coordinate-invariant form of the dielectric response was proposed recently.\cite{Schubert_2016_LST} For this analysis procedure, we augment the dielectric function form with anharmonic lattice broadening effects proposed by Berreman and Unterwald\cite{Berreman68}, and Lowndes\cite{Lowndes70} onto the coordinate-invariant generalization of the dielectric function proposed by Schubert.\cite{Schubert_2016_LST} In contrast to our previous report on $\beta$-Ga$_2$O$_3$\cite{Schubert_2016}, we do not observe the effects of free charge carriers in CdWO$_4$, and hence their contributions to the dielectric response, needed for accurate analysis of conductive, monoclinic materials such as $\beta$-Ga$_2$O$_3$, are ignored in this work. The phonon mode parameters and static and high frequency dielectric constants  obtained from our ellipsometry analysis are compared to results of density functional theory (DFT) calculations. We observe by experiment all DFT predicted modes, and all parameters including phonon mode eigenvector orientations are in excellent agreement between theory and experiment.

\section{Theory}

\subsection{Symmetry}

The cadmium tungstate unit cell contains two cadmium atoms, two tungsten atoms, and eight oxygen atoms. The lattice constants of wolframite structure CdWO$_4$ are $a=5.026$ \AA, $b=5.078$ \AA, and $c=5.867$ \AA, and the monoclinic angle is $\beta= 91.47$~\cite{Daturi_1997} (Fig.~\ref{fig:CdWO4cell}). CdWO$_4$ possesses 33 normal modes of vibration with the irreducible representation for acoustical and optical zone center modes: $\Gamma=8A_g + 10B_g + 7A_u + 8B_u$, where $A_u$ and $B_u$ modes are active at mid-infrared and far-infrared wavelengths. The phonon displacement of $A_u$ modes is parallel to the crystal $\mathbf{b}$ direction, while the phonon displacement for $B_u$ modes is parallel to the $\mathbf{a}-\mathbf{c}$ crystal plane. All modes split into transverse optical (TO) and longitudinal optical (LO) phonons.

\begin{figure*}[!hbt]
  \begin{center}
  \includegraphics[width=0.9\linewidth,natwidth=3835,natheight=2104]{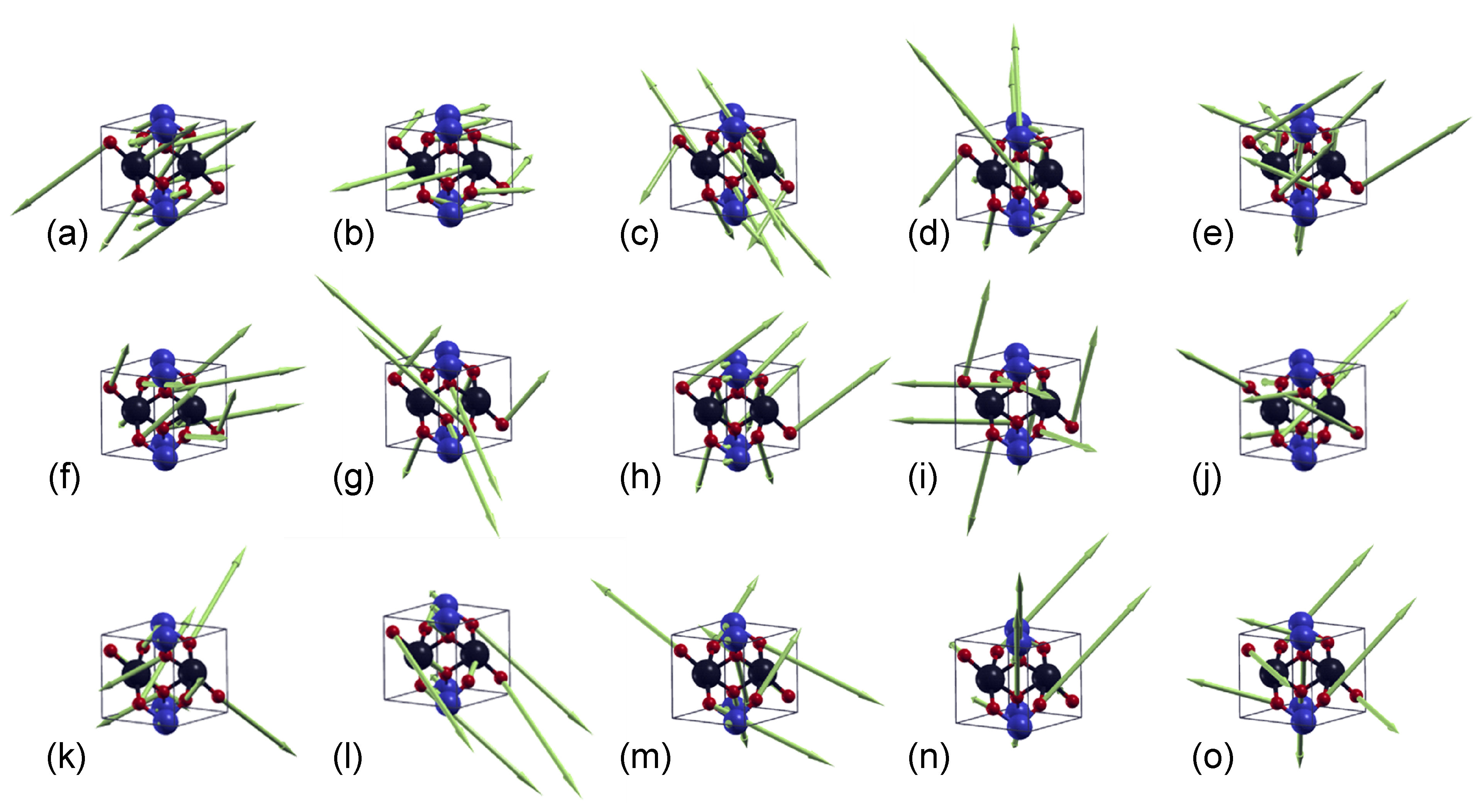}
   \caption{Renderings of TO phonon modes in CdWO$_4$ with $A_u$ (b: $A_u(7)$, g: $A_u(6)$, h: $A_u(5)$, i: $A_u(4)$, k: $A_u(3)$, m: $A_u(2)$, o: $A_u(1)$) and $B_u$ symmetry (a: $B_u(8)$, c: $B_u(7)$, d: $B_u(6)$, e: $B_u(5)$, f: $B_u(4)$, j: $B_u(3)$, l: $B_u(2)$, n: $B_u(1)$). The respective phonon mode frequency parameters calculated using Quantum Espresso are given in Tab.~\ref{tab:TOAuBuQE}.}
    \label{fig:phononrendering}
  \end{center}
\end{figure*}

\subsection{Density Functional Theory}

Theoretical calculations of long wavelength active $\Gamma$-point phonon frequencies were performed by plane wave DFT using Quantum ESPRESSO (QE)~\cite{qe}. We used the exchange correlation functional of Perdew and Zunger (PZ).~\cite{PerdewPRB1981} We employ Optimized Norm-Conserving Vanderbilt (ONCV) scalar-relativistic pseudopotentials,~\cite{Hamann2013} which we generated for the PZ functional using the code ONCVPSP\cite{ONCVPSP} with the optimized parameters of the SG15 distribution of pseudopotentials.\cite{SG15} These pseudopotentials include 20 valence states for cadmium.\cite{CdPseudo} A crystal cell of CdWO$_4$ consisting of two chemical units, with initial parameters for the cell and atom coordinates taken from Ref. ~\onlinecite{Dahlborg1999} was first relaxed to force levels less than $10^{-5}$ Ry/Bohr. A regular shifted $4\times4\times4$ Monkhorst-Pack grid was used for sampling of the Brillouin Zone~\cite{MonkhorstPRBGRID}. A convergence threshold of $1\times10^{-12}$ was used to reach self consistency with a large electronic wavefunction cutoff of 100 Ry. The fully relaxed structure was then used for the calculation of phonon modes.

\subsection{Dielectric Function Tensor Properties}
\label{sec:DFTensorModel}

\subsubsection{Transverse and longitudinal phonon modes}

From the frequency dependence of a general, linear dielectric function tensor, two mutually exclusive and characteristic sets of eigenmodes can be unambiguously defined. One set pertains to frequencies at which dielectric resonance occurs for electric fields along directions $\mathbf{\hat{e}}_{l}$. These are the so-called transverse optical (TO) modes whose eigendielectric displacement unit vectors are then $\mathbf{\hat{e}}_{l}=\mathbf{\hat{e}}_{\stext{TO},l}$. Likewise, a second set of frequencies pertains to situations when the dielectric loss approaches infinity for electric fields along directions $\mathbf{\hat{e}}_{l}$. These are the so-called longitudinal optical (LO) modes whose eigendielectric displacement unit vectors are then $\mathbf{\hat{e}}_{l}=\mathbf{\hat{e}}_{\stext{LO},l}$. This can be expressed by the following equations

\begin{subequations}\label{eq:TOLOvectors}
\begin{align}
|\det\{ \varepsilon(\omega=\omega_{\stext{TO},l})\}| &\rightarrow \infty, \\
|\det\{ \varepsilon^{-1}(\omega=\omega_{\stext{LO},l})\}| &\rightarrow \infty, \\
\varepsilon^{-1}(\omega=\omega_{\stext{TO},l})\mathbf{\hat{e}}_{\stext{TO},l} &=0,\\
\varepsilon(\omega=\omega_{\stext{LO},l})\mathbf{\hat{e}}_{\stext{LO},l} &=0.
\end{align}
\end{subequations}

\noindent where $|\zeta|$ denotes the absolute value of a complex number $\zeta$. At this point, $l$ is an index which merely addresses the occurrence of multiple such frequencies in either or both of the sets. Note that as a consequence of Eqs.~\ref{eq:TOLOvectors}, the eigendisplacement directions of TO and LO modes with a common frequency must be perpendicular to each other, regardless of crystal symmetry.

\subsubsection{The eigendielectric displacement vector summation approach}

It was shown previously that the tensor elements of $\varepsilon$ due to long wavelength active phonon modes in materials with any crystal symmetry can be obtained from an eigendielectric displacement vector summation approach. In this approach, contributions to the anisotropic dielectric polarizability from individual, eigendielectric displacements (dielectric resonances) with unit vector $\mathbf{\hat{e}}_l$ are added to a high-frequency, frequency-independent tensor, $\varepsilon_\infty$, which is thought to originate from the sum of all dielectric eigendielectric displacement processes at much shorter wavelengths than all phonon modes\cite{Schubert_2016,Schubert_2016_LST}

\begin{equation}\label{eq:epssum}
\varepsilon=\varepsilon_\infty+\sum^{N}_{l=1}\varrho_{l}(\mathbf{\hat{e}}_l\otimes\mathbf{\hat{e}}_l),
\end{equation}

\noindent where $\otimes$ is the dyadic product. Functions $\varrho_{l}$ describe the frequency responses of each of the $l=1\dots N$ eigendielectric displacement modes.\cite{footnotePRBalyssa1} Functions $\varrho_{l}$ must satisfy causality and energy conservation requirements, i.e., the Kramers-Kronig (KK) integral relations and $Im \{ \varrho_{l} \} \ge 0, \forall$ $\omega \ge 0$, $1,\dots l, \dots N$ conditions.~\cite{Dressel_2002,Jackson75}

\subsubsection{The Lorentz oscillator model}

The energy (frequency) dependent contribution to the long wavelength polarization response of an uncoupled electric dipole charge oscillation is commonly described using a KK consistent Lorentz oscillator function with harmonic broadening~\cite{SchubertIRSEBook_2004,HumlicekHOE}

\begin{equation}\label{eq:HLO}
\varrho_{l} \left(\omega\right)=\frac{A_{l}}{\omega^2_{0,l}-\omega^2-i\omega\gamma_{0,l}},
\end{equation}

or anharmonic broadening

\begin{equation}\label{eq:AHLO}
\varrho_{l} \left(\omega\right)=\frac{A_{l}-i\Gamma_{l}\omega}{\omega^2_{0,l}-\omega^2-i\omega\gamma_{0,l}},
\end{equation}

\noindent where $A_{l}$, $\omega_{0,l}$, $\gamma_{0,l}$, and $\Gamma_l$ denote amplitude, resonance frequency, harmonic broadening, and anharmonic broadening parameter of mode $l$, respectively, $\omega$ is the frequency of the driving electromagnetic field, and $i^2=-1$ is the imaginary unit. The assumption that functions $\varrho_{l}$ can be described by harmonic oscillators renders the eigendielectric displacement vector summation approach of Eq.~\ref{eq:epssum} equivalent to the result of the microscopic description of the long wavelength lattice vibrations given by Born and Huang in the so-called harmonic approximation.~\cite{Born54} In the harmonic approximation the interatomic forces are considered constant and the equations of motion are determined by harmonic potentials.\cite{VFSbook}

From Eqs.~\ref{eq:TOLOvectors}-\ref{eq:AHLO} it follows that $\mathbf{\hat{e}}_l=\mathbf{\hat{e}}_{\stext{TO},l}$, and $\omega_{0,l}=\omega_{\stext{TO},l}$. The ad-hoc parameter $\Gamma_l$ introduced in Eq.~\ref{eq:AHLO} can be shown to be directly related to the LO mode broadening parameter $\gamma_{\stext{LO},l}$ introduced previously to account for anharmonic phonon coupling in materials with orthorhombic and higher symmetries, which is discussed below.

\subsubsection{The coordinate-invariant generalized dielectric function}

The determinant of the dielectric function tensor can be expressed by the following frequency-dependent coordinate-invariant form, regardless of crystal symmetry\cite{Schubert_2016,Schubert_2016_LST}

\begin{equation}\label{eq:general-eps}
det\{\varepsilon(\omega)\}=det\{\varepsilon_\infty\}\prod_{l=1}^{N}\frac{\omega^2_{\stext{LO},l}-\omega^2}{\omega^2_{\stext{TO},l}-\omega^2}.
\end{equation}

\subsubsection{The Berreman-Unterwald-Lowndes factorized form}

The right side of Eq.~\ref{eq:general-eps} is form equivalent to the so-called factorized form of the dielectric function for long wavelength-active phonon modes described by Berreman and Unterwald\cite{Berreman68}, and Lowndes.\cite{Lowndes70} The Berreman-Unterwald-Lowndes (BUL) factorized form is convenient for derivation of TO and LO mode frequencies from the dielectric function of materials with multiple phonon modes. In the derivation of the BUL factorized form, however, it was assumed that the displacement directions of all contributing phonon modes must be parallel. Hence, in its original implementation, the application of the  BUL factorized form is limited to materials with orthorhombic, hexagonal, tetragonal, trigonal, and cubic crystal symmetries. Schubert recently suggested Eq.~\ref{eq:general-eps} as generalization of the BUL form applicable to materials regardless of crystal symmetry.\cite{Schubert_2016_LST}

\subsubsection{The generalized dielectric function with anharmonic broadening}

The introduction of broadening by permitting for parameters $\gamma_{\stext{TO},l}$ and $\Gamma_l$ in Eqs.~\ref{eq:HLO}, and ~\ref{eq:AHLO} can be shown to modify Eq.~\ref{eq:general-eps} into the following form

\begin{equation}\label{eq:general-eps-broaded}
det\{\varepsilon(\omega)\}=det\{\varepsilon_\infty\}\prod_{l=1}^{N}\frac{\omega^2_{\stext{LO},l}-\omega^2-i\omega\gamma_{\stext{LO},l}}{\omega^2_{\stext{TO},l}-\omega^2-i\omega\gamma_{\stext{TO},l}},
\end{equation}

\noindent where $\gamma_{\stext{LO},l}$ is the broadening parameter for the LO frequency $\omega_{\stext{LO},l}$. A similar augmentation was suggested by Gervais and Periou for the BUL factorized form identifying $\gamma_{\stext{LO},l}$ as independent model parameters to account for a life-time broadening mechanisms of LO modes separate from that of TO modes.\cite{Gervais74} Sometimes referred to as ``4-parameter semi quantum'' (FPSQ) model, the approach by Gervais and Periou  allowing for separate TO and LO mode broadening parameters, $\gamma_{\stext{TO},l}$ and $\gamma_{\stext{LO},l}$, respectively, provided accurate description of effects of anharmonic phonon mode coupling in anisotropic, multiple mode materials with non-cubic crystal symmetry, for example, in tetragonal (rutile) TiO$_2$,\cite{Gervais74,SchoecheJAP2013TiO2} hexagonal (corrundum) Al$_2$O$_3$,\cite{SchubertPRB61_2000} and orthorhombic (stibnite) Sb$_2$S$_3$.\cite{Schubert03g} In this work, we suggest use of Eq.~\ref{eq:general-eps-broaded} to accurately match the experimentally observed lineshapes and to determine frequencies of TO and LO modes, and thereby to account for effects of phonon mode anharmonicity in monoclinic CdWO$_4$.

\subsubsection{Schubert-Tiwald-Herzinger broadening condition}

The following condition holds for the TO and LO mode broadening parameters within a BUL form\cite{SchubertPRB61_2000}

\begin{subequations}\label{eq:STHcondition}
\begin{align}
0 &< Im\left\{\prod_{l=1}^{N}\frac{\omega^2_{\stext{LO},l}-\omega^2-i\omega\gamma_{\stext{LO},l}}{\omega^2_{\stext{TO},l}-\omega^2-i\omega\gamma_{\stext{TO},l}} \right\},\\
&\updownarrow\\
&0 < \sum\limits_{l=1}^N  {\left({\gamma _{\mathrm{LO},l} - \gamma _{\mathrm{TO},l} } \right)}.
\end{align}
\end{subequations}

This condition is valid for the dielectric function along high-symmetry Cartesian axes for orthorhombic, hexagonal, tetragonal, trigonal, and cubic crystal symmetry in materials with multiple phonon mode bands. For monoclinic materials it is valid for the dielectric function for polarizations along crystal axis \textbf{b}. Its validity for Eq.~\ref{eq:general-eps-broaded} has not been shown yet, also not for the conceptual expansion for triclinic materials (Eq.~14 in Ref.\onlinecite{Schubert_2016_LST}). However, we test the condition for the \textbf{a-c} plane in this work.

\subsubsection{Generalized Lyddane-Sachs-Teller relation}

A coordinate-invariant generalization of the Lyddane-Sachs-Teller (LST) relation\cite{Lyddane41} for arbitrary crystal symmetries was recently derived by Schubert (S-LST).\cite{Schubert_2016_LST} The S-LST relation follows immediately from Eq.~\ref{eq:general-eps-broaded} setting $\omega$ to zero

\begin{equation}\label{eq:SLST}
\frac{det\{\varepsilon \left(\omega=0 \right)\}}{det\{\varepsilon_{\infty}\}}=\prod^{N}_{l=1}\left(\frac{\omega_{\stext{LO},l}}{\omega_{\stext{TO},l}}\right)^2,
\end{equation}

\noindent and which was found valid for monoclinic $\beta$-Ga$_2$O$_3$.\cite{Schubert_2016} We investigate the validity of the S-LST relation for monoclinic CdWO$_4$ with our experimental results obtained in this work.

\subsubsection{CdWO$_4$ dielectric tensor model}\label{dtmsection}

We align unit cell axes $\mathbf{b}$ and $\mathbf{a}$ with $-z$ and $x$, respectively, and $\mathbf{c}$ is within the ($x$-$y$) plane. We introduce vector $\mathbf{c^{\star}}$ parallel to $y$ for convenience, and we obtain $\mathbf{a}$, $\mathbf{c^{\star}}$, $-\mathbf{b}$ as a pseudo orthorhombic system (Fig.~\ref{fig:CdWO4cell}). Seven modes with $A_u$ symmetry are polarized along $\mathbf{b}$ only. Eight modes with $B_u$ symmetry are polarized within the $\mathbf{a}$ - $\mathbf{c}$ plane. For CdWO$_4$, the dielectric tensor elements are then obtained as follows

\begin{subequations}\label{eq:epsmonoall}
\begin{align}
\varepsilon_{xx} &= \varepsilon_{\infty,xx}+\sum^{8}_{l=1}\varrho^{B_u}_{l}\cos^2\alpha_j,\\
\varepsilon_{xy} &= \varepsilon_{\infty,xy}+\sum^{8}_{l=1}\varrho^{B_u}_{l}\sin\alpha_j\cos\alpha_j,\\
\varepsilon_{yy} &= \varepsilon_{\infty,yy}+\sum^{8}_{l=1}\varrho^{B_u}_{l}\sin^2\alpha_j,\\
\varepsilon_{zz} &= \varepsilon_{\infty,zz}+\sum^{7}_{l=1}\varrho^{A_u}_{l},\\
\varepsilon_{xz} &= \varepsilon_{zx} = \varepsilon_{zy}= \varepsilon_{yz}=0,
\end{align}
\end{subequations}

\noindent where $X=A_u, B_u$ indicate functions $\varrho_l^X$ for long wavelength active modes with $A_u$ and $B_u$ symmetry, respectively. The angle $\alpha_l$ denotes the orientation of the eigendielectric displacement vectors with $B_u$ symmetry relative to axis $\mathbf{a}$. Note that the eigendielectric displacement vectors with $A_u$ symmetry are all parallel to axis $\mathbf{b}$, and hence do not appear as variables in Eqs.~\ref{eq:epsmonoall}.

\subsubsection{Phonon mode parameter determination}

The spectral dependence of the CdWO$_4$ dielectric function tensor, obtained here by generalized ellipsometry measurements, is performed in two stages. The first stage does not involve assumptions about a physical lineshape model. The second stage applies the eigendielectric displacement vector summation approach described above.

Stage 1, according to Eqs.~\ref{eq:TOLOvectors}, the elements of experimentally determined $\varepsilon$ and $\varepsilon^{-1}$ are plotted versus wavelength, and $\omega_{\stext{TO},l}$ and $\omega_{\stext{LO},l}$, are determined from extrema in $\varepsilon$ and $\varepsilon^{-1}$, respectively. Eigenvectors $\hat{\mathbf{e}}_{\stext{TO},l}$ and $\hat{\mathbf{e}}_{\stext{LO},l}$ can be estimated by solving Eq.~\ref{eq:TOLOvectors}(c) and ~\ref{eq:TOLOvectors}(d), respectively.

Stage 2, step (i): Eqs.~\ref{eq:epsmonoall} are used to match simultaneously all elements of the experimentally determined tensors $\varepsilon$ and $\varepsilon^{-1}$. As a result, $\varepsilon_{\infty}$ and eigenvector, amplitude, frequency, and broadening parameters for all TO modes are obtained. Step (ii): ($B_u$ symmetry) The generalized dielectric function (Eq.~\ref{eq:general-eps-broaded}) is used to determine the LO mode frequency and broadening parameters. All other parameters in Eq.~\ref{eq:general-eps-broaded} are taken from step (i). The eigenvectors $\hat{\mathbf{e}}_{\stext{LO},l}$ are calculated by solving Eq.~\ref{eq:TOLOvectors}(d). ($A_u$ symmetry) The BUL form is used to parameterize $\varepsilon_{zz}$ and $-\varepsilon^{-1}_{zz}$ in order to determine the LO mode frequency and broadening parameters.

\subsection{Generalized Ellipsometry}

Generalized ellipsometry is a versatile concept for analysis of optical properties of generally anisotropic materials in bulk as well as in multiple-layer stacks.\cite{SchubertPRB61_2000,SchoecheJAP2013TiO2,Schubert03g,DresselOE2008pentacene,Schubert03c,Ashkenov03,KasicPRB62_2000,KasicPRB65_2002,DarakchievaAPL84_2004,DarakchievaPRB2004AlN,DarakchievaPRB2005AlGaNSL,DarakchievaAPL2009InNe,DarakchievaAPL2009InN,DarakchievaAPL2010InN,DarakchievaPRB2014InNmix,DarakchievaJAP2014InNMg,HofmannTHzGLADChapter2013} A multiple sample, multiple azimuth, and multiple angle of incidence approach is required for monoclinic CdWO$_4$, following the same approach used previously for monoclinic $\beta$-Ga$_2$O$_3$.\cite{Schubert_2016} Multiple, single crystalline samples cut under different angles from the same crystal must be investigated and analyzed simultaneously.

\subsubsection{Mueller matrix formalism}

In generalized ellipsometry, either the Jones or the Mueller matrix formalism can be used to describe the interaction of electromagnetic plane waves with layered samples~\cite{SchubertIRSEBook_2004,Thompkins_2004,HumlicekHOE,Azzam95,Fujiwara_2007}. In the Mueller matrix formalism, real-valued Mueller matrix elements connect the Stokes parameters of the electromagnetic plane waves before and after sample interaction

\begin{equation}
\left( {{\begin{array}{*{20}c}
 {S_{0} } \hfill \\ {S_{1} } \hfill \\  {S_{2} } \hfill \\  {S_{3} } \hfill \\
\end{array} }} \right)_{\mathrm{output}} =
\left( {{\begin{array}{*{20}c}
 {M_{11} } \hfill & {M_{12} } \hfill \ {M_{13} } \hfill & {M_{14} } \hfill \\
 {M_{21} } \hfill & {M_{22} } \hfill \ {M_{23} } \hfill & {M_{24} } \hfill \\
 {M_{31} } \hfill & {M_{32} } \hfill \ {M_{33} } \hfill & {M_{34} } \hfill \\
 {M_{41} } \hfill & {M_{42} } \hfill \ {M_{43} } \hfill & {M_{44} } \hfill \\
\end{array} }} \right)
\left( {{\begin{array}{*{20}c}
 {S_{0} } \hfill \\ {S_{1} } \hfill \\  {S_{2} } \hfill \\  {S_{3} } \hfill \\
\end{array} }} \right)_{\mathrm{input}}.
\end{equation}
with the Stokes vector components defined by $S_{0}=I_{p}+I_{s}$, $S_{1}=I_{p} - I_{s}$, $S_{2}=I_{45}-I_{ -45}$, $S_{3}=I_{\sigma + }-I_{\sigma - }$, where $I_{p}$, $I_{s}$, $ I_{45}$, $I_{-45}$, $I_{\sigma + }$, and $I_{\sigma - }$denote the intensities for the $p$-, $s$-, +45$^{\circ}$, -45$^{\circ}$, right handed, and left handed circularly polarized light components, respectively~\cite{Fujiwara_2007}. The Mueller matrix renders the optical sample properties at a given angle of incidence and sample azimuth, and data measured must be analyzed through a best match model calculation procedure in order to extract relevant physical parameters.\cite{JellisonHOE_2004,Aspnes98}

\subsubsection{Model analysis}

The $4\times 4$ matrix formalism is used to calculate the Mueller matrix. We apply the half-infinite two-phase model, where ambient (air) and monoclinic CdWO$_4$ render the two half infinite mediums separated by the plane at the surface of the single crystal. The formalism has been detailed extensively.~\cite{Schubert96,Schubert03a,SchubertIRSEBook_2004,Schubert04,Fujiwara_2007} The only free parameters in this approach are the elements of the dielectric function tensor of the material with monoclinic crystal symmetry, and the angle of incidence. The latter is set by the instrumentation. The wavelength only enters this model explicitly when the dielectric function tensor elements are expressed by wavelength dependent model functions. This fact permits the determination of the dielectric function tensor elements in the so-called wavelength-by-wavelength model analysis approach.

\subsubsection{Wavelength-by-wavelength analysis}

Two coordinate systems must be established such that one that is tied to the instrument and another is tied to the crystallographic sample description. The system tied to the instrument is the system in which the dielectric function tensor must be cast into for the 4$\times$4 matrix algorithm. We chose both coordinate systems to be Cartesian. The sample normal defines the laboratory coordinate system's $\hat{z}$ axis, which points into the surface of the sample.\cite{Schubert96,Schubert_2016} The sample surface then defines the laboratory coordinate system's $\hat{x}$ - $\hat{y}$ plane. The sample surface is at the origin of the coordinate system. The plane of incidence is the $\hat{x}$ - $\hat{z}$ plane. Note that the system ($\hat{x}$, $\hat{y}$, $\hat{z}$) is defined by the ellipsometer instrumentation through the plane of incidence and the sample holder. One may refer to this system as the laboratory coordinate system. The system ($x$, $y$, $z$) is fixed by our choice to the specific orientation of the CdWO$_4$ crystal axes, \textbf{a}, \textbf{b}, and \textbf{c} as shown in Fig.~\ref{fig:CdWO4cell} with vector $\mathbf{c^{\star}}$ defined for convenience perpendicular to \textbf{a-b} plane. One may refer to system ($x$, $y$, $z$) as our CdWO$_4$ system. Then, the full dielectric tensor in the 4$\times$4 matrix algorithm is
\begin{equation}\label{eq:monoclinicepsDC}
\boldsymbol{\varepsilon}= \left(
\begin{array}{ccc}
\varepsilon_{xx} & \varepsilon_{xy} & 0 \\
\varepsilon_{xy} & \varepsilon_{yy} & 0 \\
0          & 0 & \varepsilon_{zz}

\end{array}\right),
\end{equation}
with elements obtained by setting $\varepsilon_{xx}, \varepsilon_{xy}, \varepsilon_{yy}$, and $\varepsilon_{zz}$ as unknown parameters. Then, according to the crystallographic surface orientation of a given sample, and according to its azimuth orientation relative to the plane of incidence, a Euler angle rotation is applied to $\varepsilon$. The sample azimuth angle, typically termed $\varphi$, is defined by a certain in plane rotation with respect to the sample normal. The sample azimuth angle describes the mathematical rotation that a model dielectric function tensor of a specific sample must make when comparing calculated data with measured data from one or multiple samples taken at multiple, different azimuth positions.

As first step in data analysis, all ellipsometry data were analyzed using a wavelength-by-wavelength approach. Model calculated Mueller matrix data were compared to experimental Mueller matrix data, and dielectric tensor values were varied until best match was obtained. This is done by minimizing the mean square error ($\chi^2$) function which is weighed to estimated experimental errors ($\sigma$) determined by the instrument for each data point~\cite{SchubertPRB61_2000,Schubert03h,SchubertIRSEBook_2004,SchubertADP15_2006,SchoecheJAP2013TiO2}. The error bars on the best match model calculated tensor parameters then refer to the usual 90\% confidence interval. All data obtained at the same wavenumber from multiple samples, multiple azimuth angles, and multiple angles of incidence are included (polyfit) and one set of complex values $\varepsilon_{xx}, \varepsilon_{xy}, \varepsilon_{yy}$, and $\varepsilon_{zz}$ is obtained. This procedure is simultaneously and independently performed for all wavelengths. In addition, each sample requires one set of 3 independent Euler angle parameters, each set addressing the orientation of axes $\mathbf{a}$, $\mathbf{b}$, $\mathbf{c^{\star}}$ at the first azimuth position where data were acquired.

\subsubsection{Model dielectric function analysis}
\label{sec:Ellipsometryanalyses}

A second analysis step is performed by minimizing the difference between the wavelength-by-wavelength extracted $\varepsilon_{xx}, \varepsilon_{xy}, \varepsilon_{yy}$, and $\varepsilon_{zz}$ spectra and those calculated by Eqs.~(\ref{eq:epsmonoall}). All model parameters were varied until calculated and experimental data matched as close as possible (best match model). For the second analysis step, the numerical uncertainty limits of the 90\% confidence interval from the first regression were used as ``experimental'' errors $\sigma$ for the wavelength-by-wavelength determined $\varepsilon_{xx}, \varepsilon_{xy}, \varepsilon_{yy}$, and $\varepsilon_{zz}$ spectra. A similar approach was described, for example, in Refs.~\onlinecite{SchubertPRB61_2000,HofmannPRB66_2002,SchubertIRSEBook_2004,SchoecheJAP2013TiO2,Schubert_2016}. All best match model calculations were performed using the software package WVASE$^{32}$ (J.~A.~Woollam~Co.,~Inc.).

\section{Experiment}

Two single crystal samples of CdWO$_4$ were fabricated by slicing from a bulk crystal with (001) and (010) surface orientation. Both samples were then double side polished. The substrate dimensions are 10mm$\times$10mm$\times$0.5mm for the (001) crystal and 10mm$\times$10mm$\times$0.2mm for the (010) crystal.

Mid-infrared  (IR) and far-infrared (FIR) generalized spectroscopic ellipsometry (GSE) were performed at room temperature on both samples. The IR-GSE measurements were performed on a rotating compensator infrared ellipsometer (J.~A.~Woollam Co., Inc.) in the spectral range from 333 -- 1500 cm$^{-1}$ with a spectral resolution of 2 cm$^{-1}$. The FIR-GSE measurements were performed on an in-house built rotating polarizer rotating analyzer far-infrared ellipsometer in the spectral range from 50 -- 500 cm$^{-1}$ with an average spectral resolution of 1 cm$^{-1}$.~\cite{KuehneRSI2014} All GSE measurements were performed at 50$^\circ$, 60$^\circ$, and 70$^\circ$ angles of incidence. All measurements are reported in terms of Mueller matrix elements, which are normalized to element $M_{11}$. The IR instrument determines the normalized Mueller matrix elements except for those in the forth row. Note that due to the lack of a compensator for the FIR range in this work, neither element in the fourth row nor fourth column of the Mueller matrix is obtained with our FIR ellipsometer. Data were acquired at 8 in-plane azimuth rotations for each sample. The azimuth positions were adjusted by progressive, counterclockwise steps of 45$^{\circ}$.

\section{Results and Discussion}
\subsection{DFT Phonon Calculations}
\begin{table*}[!hbt]
\setlength{\tabcolsep}{0.75pt}
\caption{Phonon mode parameters for $A_u$ and $B_u$ modes of CdWO$_4$ obtained from DFT calculations using Quantum Espresso. Renderings of displacements are shown in Fig.~\ref{fig:phononrendering}. }
\begin{center}
\begin{tabular}{lcccccccccccccccc}
    \noalign{\bigskip} \hline \hline
		& &$X=B_u$ & & & & & & & &$X=A_u$ & & & & & & \\
    \cline{3-10}\cline{11-17}
    Parameter& &k=1&2&3&4&5&6&7&8&k=1&2&3&4&5&6&7\\
    \noalign{\smallskip} \hline
    $A_{k}^{X}$ [(\textit{e}B)$^2$/2]  & this work &2.61&3.31&0.29&1.38&0.12&0.18&0.27&0.10&0.52&1.47&0.65&0.28&0.43&0.03&0.15\\
    $\omega_{\stext{TO},k}$ [cm$^{-1}$] &this work& 786.47&565.46&458.33&285.00&264.05&225.70&156.97&108.50&863.40&669.13&510.16&407.97&329.74&285.88&138.11\\
		$\alpha_{\stext{TO},k}$ [$^{\circ}$]&this work &22.9&111.5&8.3&69.9&59.8&126.8&157.2&28.3& - & - & - & - & - & - & - \\ \hline
		$\omega_{\stext{TO},k}$ [cm$^{-1}$] & Ref. \onlinecite{Lacomba-Perales_2009} & 743.6&524.2&420.9&255.2&252.9&225.9&145.0&105.6&839.1&626.8&471.4&379.4&322.1&270.1&121.5 \\
		\hline \hline
\end{tabular} \label{tab:TOAuBuQE}
\end{center}
\end{table*}

The phonon frequencies and transition dipole components were computed at the $\Gamma$-point of the Brillouin zone using density functional perturbation theory.~\cite{BaroniRMP2001DFTPhonons} The results of the phonon mode calculations for all long wavelength active modes with $A_u$ and $B_u$ symmetry are listed in Tab.~\ref{tab:TOAuBuQE}. Data listed include the TO resonance frequencies, and for modes with $B_u$ symmetry the angles of the transition dipoles relative to axis $\mathbf{a}$ within the $\mathbf{a} - \mathbf{c}$ plane. Renderings of atomic displacements for each mode were prepared using XCrysDen~\cite{XCrysDen} running under Silicon Graphics Irix 6.5, and are shown in Fig.~\ref{fig:phononrendering}. Frequencies of TO modes calculated by Lacomba-Perales \textit{et al.} (Ref. \onlinecite{Lacomba-Perales_2009}) using GGA-DFT are included in Tab. \ref{tab:TOAuBuQE} for reference. We note that data from Ref. \onlinecite{Lacomba-Perales_2009} are considerably shifted with respect to ours, while our calculated data agree very closely with our experimental results as discussed below.

\subsection{Mueller matrix analysis}

\begin{figure*}[!tbp]
  \begin{center}
        \includegraphics[width=1\linewidth,natwidth=4394,natheight=3379]{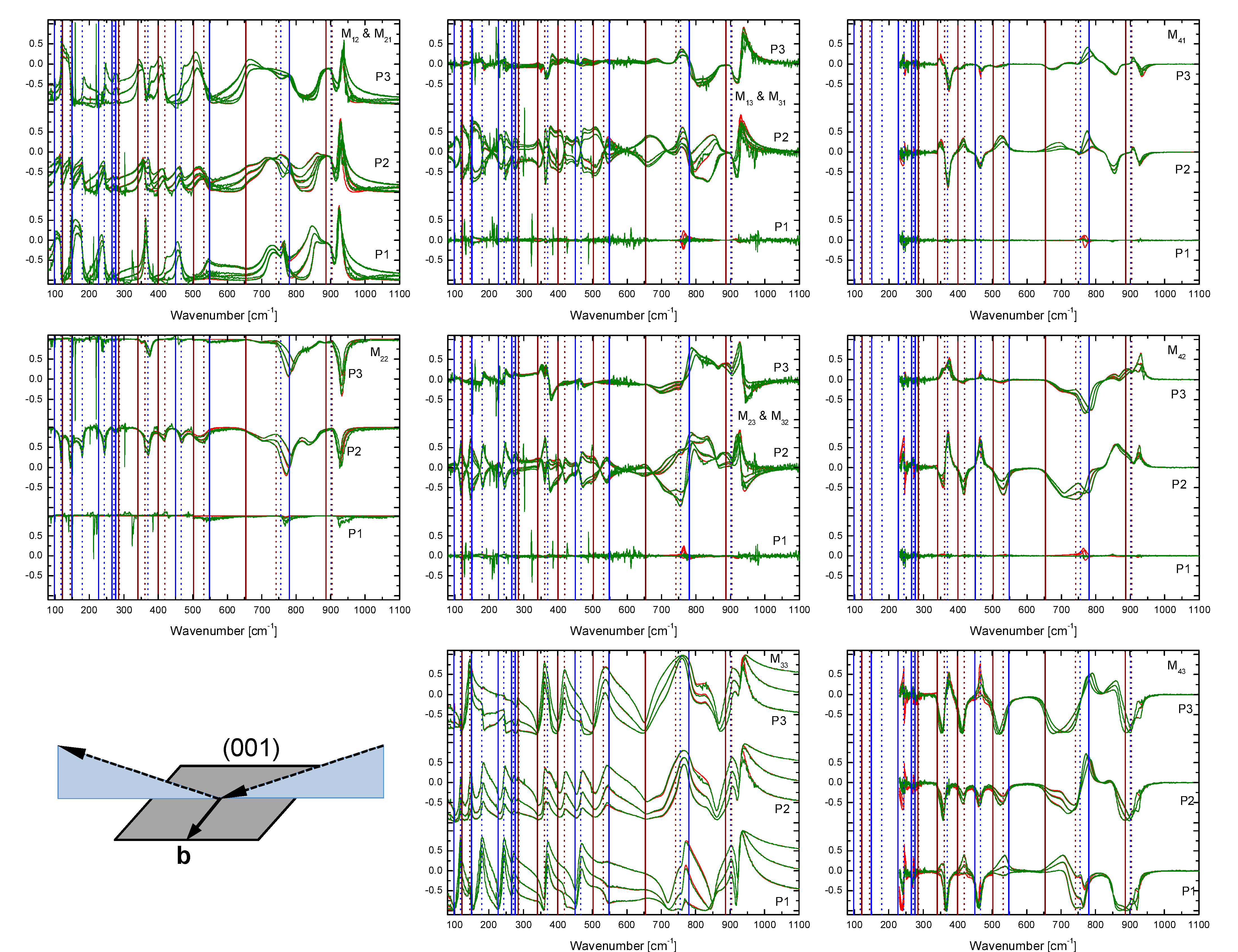}
   \caption{Experimental (dotted, green lines) and best match model calculated (solid, red lines) Mueller matrix data obtained from a (001) surface at three representative sample azimuth orientations. (P1: $\varphi=-1.3(1)^{\circ}$, P2: $\varphi=43.7(1)^{\circ}$, P3: $\varphi=88.7(1)^{\circ}$). Data were taken at three angles of incidence ($\Phi_a=50^{\circ}, 60^{\circ}, 70^{\circ}$). Equal Mueller matrix data, symmetric in their indices, are plotted within the same panels for convenience. Vertical lines indicate wavenumbers of TO (solid lines) and LO (dotted lines) modes with $B_u$ symmetry (blue) and $A_u$ symmetry (brown). Fourth column elements are only available from the IR instrument limited to approximately 333~cm$^{-1}$. Note that all elements are normalized to $M_{11}$. The remaining Euler angle parameters are $\theta=88.7(1)$ and $\psi=-1.3(1)$ consistent with the crystallographic orientation of the (001) surface. The inset depicts schematically the sample surface, the plane of incidence, and the orientation of axis $\mathbf{b}$ in P3.}
    \label{fig:exp001}
  \end{center}
\end{figure*}

\begin{figure*}[!tbp]
  \begin{center}
    \includegraphics[width=1\linewidth,natwidth=4342,natheight=3366]{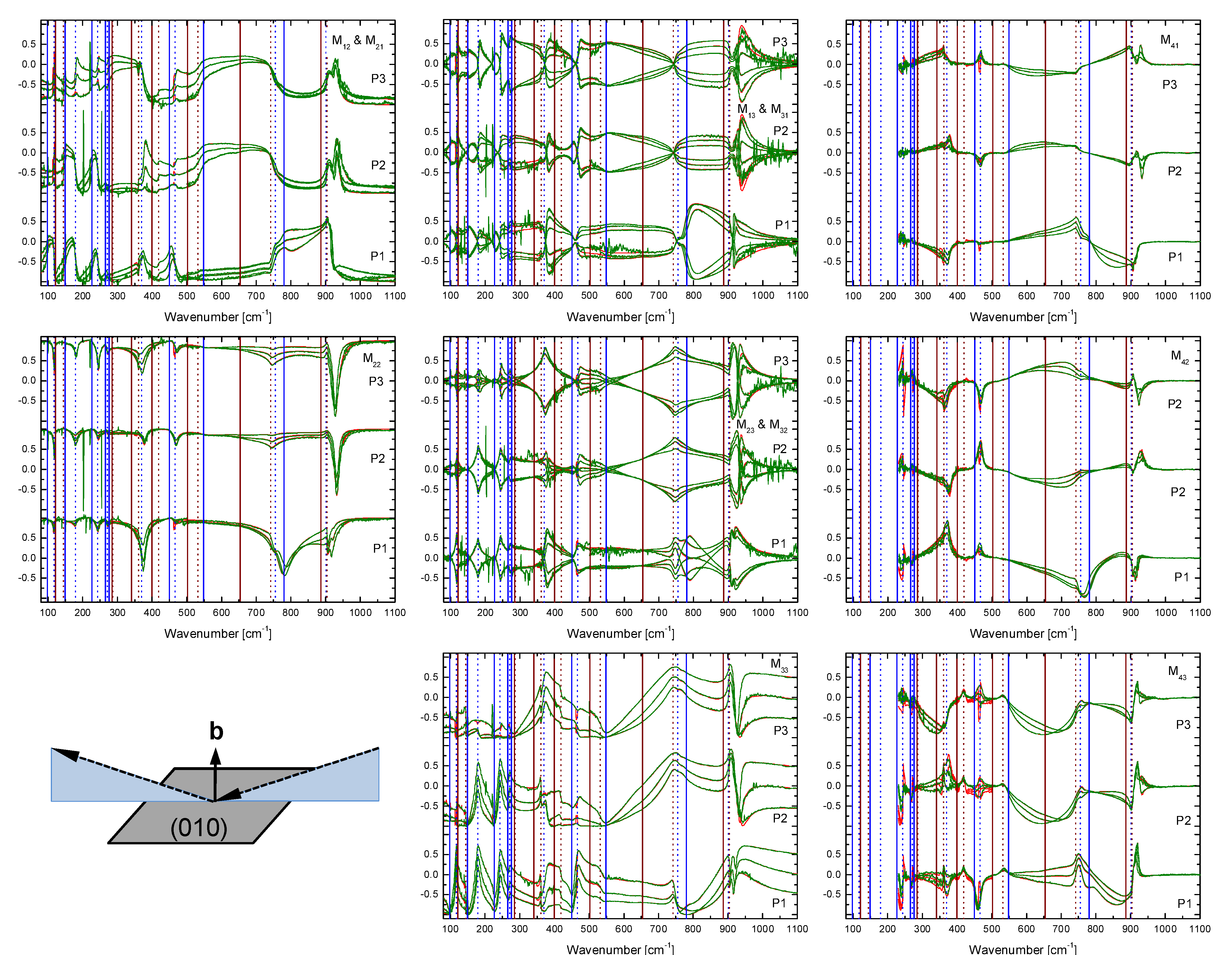}
   \caption{Same as Fig.~\ref{fig:exp010} for the (010) sample at azimuth orientation P1: $\varphi=0.5(1)^{\circ}$, P2: $\varphi=45.4(1)^{\circ}$, P3: $\varphi=90.4(1)^{\circ}$. $\theta=0.03(1)$ and $\psi=0(1)$, consistent with the crystallographic orientation of the $(010)$ surface. Note that in position P3, axis $\mathbf{b}$ which is parallel to the sample surface in this crystal cut, is aligned almost perpendicular to the plane of incidence. Hence, the monoclinic plane with $\mathbf{a}$ and $\mathbf{c}$ is nearly parallel to the plane of incidence, and as a result almost no conversion of $p$ to $s$ polarized light occurs and vice versa. As a result, the off diagonal block elements of the Mueller matrix are near zero. The inset depicts schematically the sample surface, the plane of incidence, and the orientation of axis $\mathbf{b}$, shown approximately for position P3.}
    \label{fig:exp010}
  \end{center}
\end{figure*}

Figures~\ref{fig:exp001} and~\ref{fig:exp010} depict representative experimental and best match model calculated Mueller matrix data for the $(001)$ and $(010)$ surfaces investigated in this work. Insets in Figures~\ref{fig:exp001} and~\ref{fig:exp010} show schematically axis $\mathbf{b}$ within the sample surface and perpendicular to the surface, respectively, and the plane of incidence is also indicated. Graphs depict selected data, obtained at 3 different sample azimuth orientations each $45^{\circ}$ apart. Panels with individual Mueller matrix elements are shown separately, and individual panels are arranged according to the indices of the Mueller matrix element. It is observed by experiment as well as by model calculations that all Mueller matrix elements are symmetric, i.e., $M_{ij}=M_{ji}$.
Hence, elements with $M_{ij}=M_{ji}$, i.e., from upper and lower diagonal parts of the Mueller matrix, are plotted within the same panels. Therefore, the panels represent the upper part of a $4\times4$ matrix arrangement. Because all data obtained are normalized to element $M_{11}$, and because $M_{1j}=M_{j1}$, the first column does not appear in this arrangement. The only missing element is $M_{44}$, which cannot be obtained in our current instrument configuration due to the lack of a second compensator. Data are shown for wavenumbers (frequencies) from 80 cm$^{-1}$ -- 1100 cm$^{-1}$, except for column $M_{4j}=M_{j4}$ which only contains data from approximately 250 cm$^{-1}$ -- 1100 cm$^{-1}$. All other panels show data obtained within the FIR range (80 cm$^{-1}$ -- 500 cm$^{-1}$) using our FIR instrumentation and data obtained within the IR range (500 cm$^{-1}$ -- 1100 cm$^{-1}$) using our IR instrumentation. Data from the remaining 5 azimuth orientations for each sample at which measurements were also taken are not shown for brevity.

The most notable observation from the experimental Mueller matrix data behavior is the strong anisotropy, which is reflected by the non vanishing off diagonal block elements $M_{13}$, $M_{23}$, $M_{14}$, and $M_{24}$, and the strong dependence on sample azimuth in all elements. A noticeable observation is that the off diagonal block elements in position P1 for the $(001)$ surface in Fig.~\ref{fig:exp001} are close to zero. There, axis $\mathbf{b}$ is aligned almost perpendicular to the plane of incidence. Hence, the monoclinic plane with $\mathbf{a}$ and $\mathbf{c}$ is nearly parallel to the plane of incidence, and as a result almost no conversion of $p$ to $s$ polarized light occurs and vice versa. As a result, the off diagonal block elements of the Mueller matrix are near zero. The reflected light for $s$ polarization is determined by $\varepsilon_{zz}$ alone, while the $p$ polarization receives contribution from $\varepsilon_{xx}$, $\varepsilon_{xy}$, and $\varepsilon_{yy}$, which then vary with the angle of incidence. A similar observation was made previously for a $(\bar{2}01)$ surface of monoclinic $\beta$=Ga$_2$O$_3$.\cite{Schubert_2016} While every data set (sample, position, azimuth, angle of incidence) is unique, all data sets share characteristic features at certain wavelengths. Vertical lines indicate frequencies, which further below we will identify with the frequencies of all anticipated TO and LO phonon mode mode frequencies with $A_u$ and $B_u$ symmetries. All Mueller matrix data were analyzed simultaneously during the polyfit, wavelength-by-wavelength best match model procedure. For every wavelength, up to 528 independent data points were included from the different samples, azimuth positions, and angle of incidence measurements, while only 8 independent parameters for real and imaginary parts of $\varepsilon_{xx}$, $\varepsilon_{xy}$, $\varepsilon_{yy}$, and $\varepsilon_{zz}$ were searched for. In addition, two sets of 3 wavelength independent Euler angle parameters were looked for. The results of polyfit calculation are shown in Figs.~\ref{fig:exp001} and~\ref{fig:exp010} as solid lines for the Mueller matrix elements.  We note in Figs.~\ref{fig:exp001} and~\ref{fig:exp010} the excellent agreement between measured and model calculated Mueller matrix data. Furthermore, the Euler angle parameters, given in captions of Figs.~\ref{fig:exp001} and~\ref{fig:exp010}, are in excellent agreement with the anticipated orientations of the crystallographic sample axes.

\subsection{Dielectric tensor analysis}
\begin{figure*}[!tbp]
\begin{center}
   \includegraphics[width=0.99\linewidth,natwidth=4439,natheight=3372]{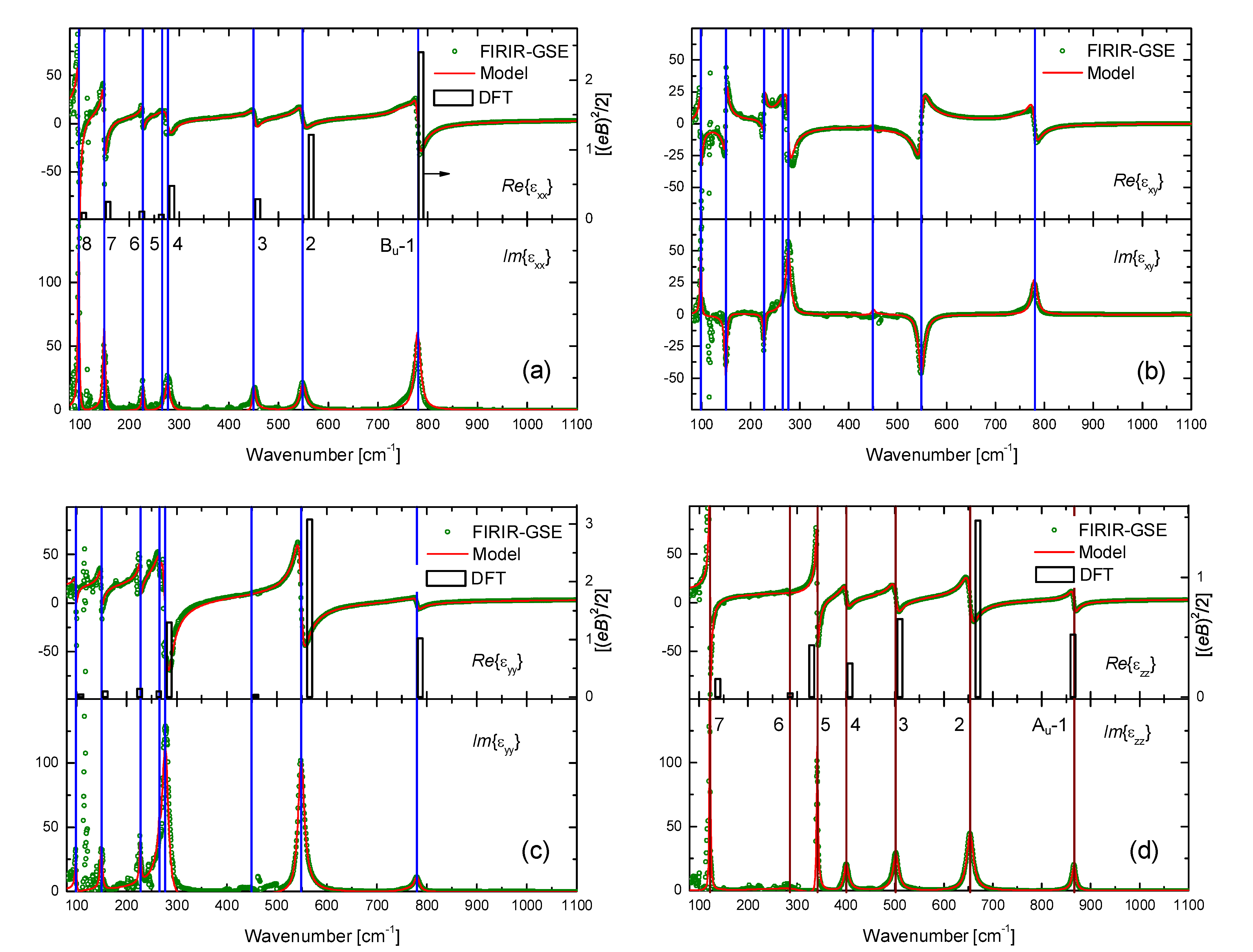}

    \caption{Dielectric function tensor element $\varepsilon_{xx}$ (a), $\varepsilon_{xy}$ (b), $\varepsilon_{yy}$ (c), and $\varepsilon_{zz}$ (d). Dotted lines (green) indicate results from wavelength by wavelength best match model regression analysis matching the  experimental Mueller matrix data shown in Figs.~\ref{fig:exp010} and~~\ref{fig:exp001}. Solid lines are obtained from best match model lineshape analysis using Eqs.~\ref{eq:epsmonoall} with Eq.~\ref{eq:AHLO}. Vertical lines in panel group [(a), (b), (c)], and in panel (d) indicate TO frequencies with $B_u$ and $A_u$ symmetry, respectively. Vertical bars in (a), (c), and (d) indicate DFT calculated long wavelength transition dipole moments  in atomic units projected onto axis $x$, $y$, and $z$, respectively.}
    \label{fig:epsilonmatrix}
  \end{center}
\end{figure*}
\begin{figure*}[!tbp]
\begin{center}
    \includegraphics[width=0.99\linewidth,natwidth=4417,natheight=3363]{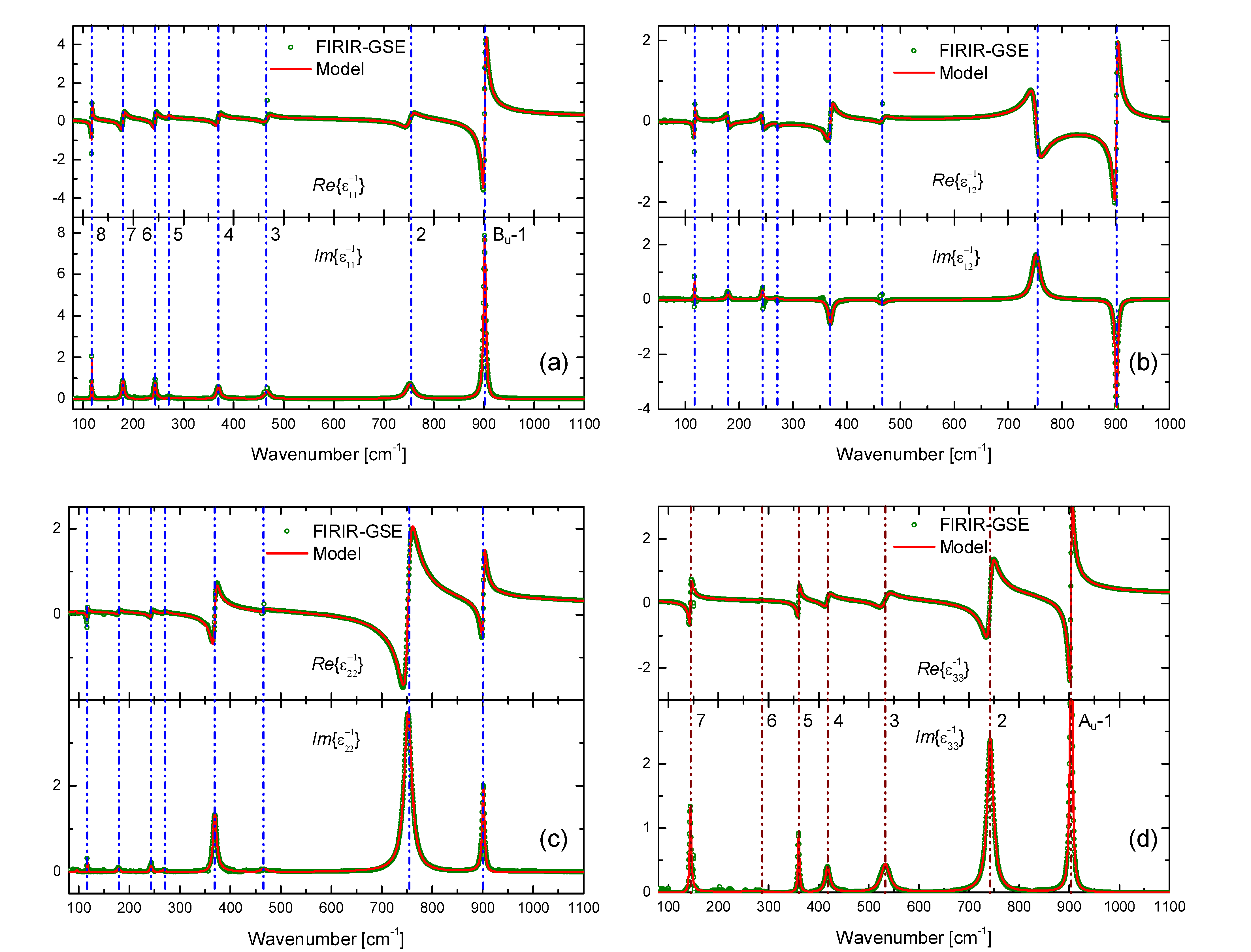}

    \caption{Same as Fig.~\ref{fig:epsilonmatrix} for the inverse dielectric tensor elements. Vertical lines in panel group [(a), (b), (c)], and in panel (d) indicate LO frequencies with $B_u$ and $A_u$ symmetry, respectively.}
    \label{fig:epsiloninversematrix}
  \end{center}
\end{figure*}
The wavelength-by-wavelength best match model dielectric function tensor data obtained during the polyfit are shown as dotted lines in Fig.~\ref{fig:epsilonmatrix} for $\varepsilon_{xx}$, $\varepsilon_{xy}$, $\varepsilon_{yy}$, and $\varepsilon_{zz}$, and in Fig.~\ref{fig:epsiloninversematrix} as dotted lines for $\varepsilon^{-1}_{xx}$, $\varepsilon^{-1}_{xy}$, $\varepsilon^{-1}_{yy}$, and $\varepsilon^{-1}_{zz}$. A detailed preview into the phonon mode properties of CdWO$_4$ is obtained here without physical lineshape analysis. In Fig.~\ref{fig:epsilonmatrix}, a set of frequencies can be identified among the tensor elements $\varepsilon_{xx}$, $\varepsilon_{xy}$, $\varepsilon_{yy}$, where their magnitudes approach large values. In particular, the imaginary parts reach large values.\cite{footnotePRBalyssa2} These frequencies are common to all elements $\varepsilon_{xx}$, $\varepsilon_{xy}$, $\varepsilon_{yy}$, and thereby reveal the frequencies of 8 TO modes with $B_u$ symmetry. The same consideration holds for $\varepsilon_{zz}$ revealing 7 LO modes with $A_u$ symmetry. The imaginary part of $\varepsilon_{xy}$ attains positive as well as negative extrema at these frequencies, and which is due to the respective eigen dielectric displacement unit vector orientation relative to axis $\mathbf{a}$. As can be inferred from Eq.~\ref{eq:epsmonoall}(b), the imaginary part of $\varepsilon_{xy}$ takes negative (positive) values when $\mathbf{\alpha}_{\stext{TO},l}$ is within $\{0 \dots -\pi \}$ ($\{0 \dots \pi \}$). Hence, $B_u$ TO modes labeled 2, 6, and 7 are oriented with negative angle towards axis $\mathbf{a}$.
A similar observation can be made in Fig.~\ref{fig:epsiloninversematrix}, where a set of frequencies can be identified among the tensor elements $\varepsilon^{-1}_{xx}$, $\varepsilon^{-1}_{xy}$, $\varepsilon^{-1}_{yy}$ when magnitudes approach large values. These frequencies are again common to all elements $\varepsilon^{-1}_{xx}$, $\varepsilon^{-1}_{xy}$, $\varepsilon^{-1}_{yy}$, and thereby reveal the frequencies of 8 LO modes with $B_u$ symmetry. The same consideration holds for $\varepsilon_{zz}$ revealing 7 LO modes with $A_u$ symmetry. The imaginary part of $\varepsilon^{-1}_{xy}$ attains positive as well as negative extrema at these frequencies, and which is due to the respective LO eigen dielectric displacement unit vector orientation relative to axis $\mathbf{a}$. We note that depicting the imaginary parts of $\varepsilon$ and $\varepsilon^{-1}$ alone would suffice to identify the phonon mode information discussed above. We further note that the inverse tensor does not contain new information, however, in this presentation the properties of the two sets of phonon modes are most conveniently visible. We finally note that up to this point no physical model lineshape model was applied.

\subsection{Phonon mode analysis}
\begin{table*}[!t]
\caption{Phonon mode parameters with $A_u$ and $B_u$ symmetries obtained from best match model analysis of tensor element spectra $\varepsilon$ and $\varepsilon^{-1}$, using anharmonic broadened Lorentz oscillator functions in Eq.~\ref{eq:AHLO}. LO mode frequency and broadening parameters are obtained from the generalized coordinate invariant form of the dielectric function proposed by Schubert.\cite{Schubert_2016_LST} The last digit, which is determined within the 90\% confidence interval, is indicated with brackets for each parameter.}
\begin{center}
\begin{tabular}{lcccccccc}
    \noalign{\bigskip} \hline \hline\noalign{\smallskip}
& $X=A_u$ & & & & & & &\\
Parameter&l=1&2&3&4&5&6&7&8\\
\hline
\noalign{\smallskip}
$\omega_{\stext{TO},l}^X$ (cm$^{-1}$)&779.5(1)&549.0(1)&450.6(2)&276.3(1)&265.2(2)&227.3(1)&149.1(1)&98.1(1)\\
$\gamma_{\stext{TO},l}^X$ (cm$^{-1}$)&15.0(1)&15.3(1)&12.5(4)&11.3(1)&12.0(4)&5.0(1)&5.7(1)&3.5(1)\\
$\alpha_{\stext{TO},l}$ ($^\circ$)&24.3(1)&-66.9(1)&180.8(8)&65.6(1)&-98.1(4)&-52.4(5)&145.1(3)&18.9(3)\\
$A_{l}^X$ (cm$^{-1}$)&908(1)&1018(1)&279(2)&645(3)&326(6)&236(1)&294(1)&236(1)\\
$\Gamma_{l}^X$ (cm$^{-1}$)&31(1)&-22(2)&-17(2)&-67(7)&88(8)&7(1)&-27(1)&70(1)\\
$\omega^X_{\stext{LO},l}$ (cm$^{-1}$)&901.4(1)&754.4(1)&466.5(1)&369.8(1)&269.1(2)&243.5(1)&180.0(1)&117.0(1)\\
$\gamma^X_{\stext{LO},l}$ (cm$^{-1}$)&5.6(1)&20.2(2)&16.6(2)&9.1(1)&12.9(4)&5.1(1)&8.0(1)&7.5(2)\\
$\alpha_{\stext{LO},l}$ ($^\circ$)&55.1&-23.6&68.7&31.5&-59.5&-66.3&-73.1&67.0\\
\noalign{\smallskip}
    \hline
    \noalign{\smallskip}
&$X=B_u$ & & & & & & &\\
&l=1&2&3&4&5&6&7&\\
    \hline\noalign{\smallskip}
$\omega_{\stext{TO},l}^X$ (cm$^{-1}$)&866.6(1)&653.7(1)&501.0(1)&400.3(1)&341.2(1)&285.5(8)&121.8(1)&\\
$\gamma_{\stext{TO},l}^X$ (cm$^{-1}$)&7.5(1)&15.8(1)&15.1(2)&10.2(2)&3.4(1)&17(1)&2.0(1)&\\
$A_{l}^X$ (cm$^{-1}$)&392(1)&679(1)&445(1)&299(1)&364(1)&93(6)&226(1)&\\
$\Gamma_k^X$ (cm$^{-1}$)&8.6(3)&14(1)&-29(1)&-24(1)&-16(1)&57(3)&-9.4(4)&\\
$\omega^X_{\stext{LO},l}$ (cm$^{-1}$)&904.0(1)&742.4(1)&532.8(1)&418.0(1)&360.2(1)&286.8(1)&144.0(1)&\\
$\gamma^X_{\stext{LO},l}$ (cm$^{-1}$)&5.1(1)&15.0(1)&19.5(2)&12.1(1)&3.5(1)&11.8(2)&3.5(1)&\\
\noalign{\smallskip}
    \hline\hline
\end{tabular} \label{tab:TOAuBuGSE}
\end{center}
\end{table*}

 \paragraph{TO modes:} Figs.~\ref{fig:epsilonmatrix} and~\ref{fig:epsiloninversematrix} depict solid lines obtained from the best match mode calculations using Eqs.~\ref{eq:epsmonoall} and the anharmonic broadened Lorentz oscillator functions in Eq.~\ref{eq:AHLO}. We find excellent match between all spectra of both tensors $\varepsilon$ and $\varepsilon^{-1}$. \cite{footnotePRBalyssa3} The best match model parameters are summarized in Tab.~\ref{tab:TOAuBuGSE}. As a result, we obtain amplitude, broadening, frequency, and eigenvector parameters for all TO modes with $A_u$ and $B_u$ symmetries. We find 8 TO mode frequencies with $B_u$ symmetry and 7 with $A_u$ symmetry. Their frequencies are indicated by vertical lines in panel group [(a), (b), (c)] and panel (d) of Fig.~\ref{fig:epsilonmatrix}, respectively, and which are identical to those observed by the extrema in the imaginary parts of the dielectric tensor components discussed above. As discussed in Sect.~\ref{sec:DFTensorModel}, element $\varepsilon_{xy}$ provides insight into the relative orientation of the unit eigen displacement vectors for each TO mode within the $\mathbf{a}$ - $\mathbf{c}$ plane. In particular, modes $B_u$-2, $B_u$-6, and $B_u$-7 reveal eigenvectors within the interval $\{ 0 \dots -\pi \}$, and cause negative imaginary resonance features in $\varepsilon_{xy}$. Accordingly, their unit eigen displacement vectors in Tab.~\ref{tab:TOAuBuGSE} reflect values larger than 90$^{\circ}$. The remaining mode unit vectors possess values between $\{ 0 \dots \pi \}$ and their resonance features in the imaginary part of $\varepsilon_{xy}$ are positive.

 Previous reports have been made of CdWO$_4$ TO mode frequencies and their symmetry assignments for \cite{Burshtein,Blasse_1975,Daturi_1997,Gabrusenoks_2001}, however, none provide a complete set of IR active modes. Due to biaxial anisotropy from the monoclinic crystal, reflectivity measurements do not provide enough information to determine directions of the TO eigenvectors. Therefore, no previously determined TO mode frequencies could be accurately compared here.

\begin{figure}[!tbp]
  \begin{center}
      \includegraphics[width=0.99\linewidth,natwidth=2096,natheight=3441]{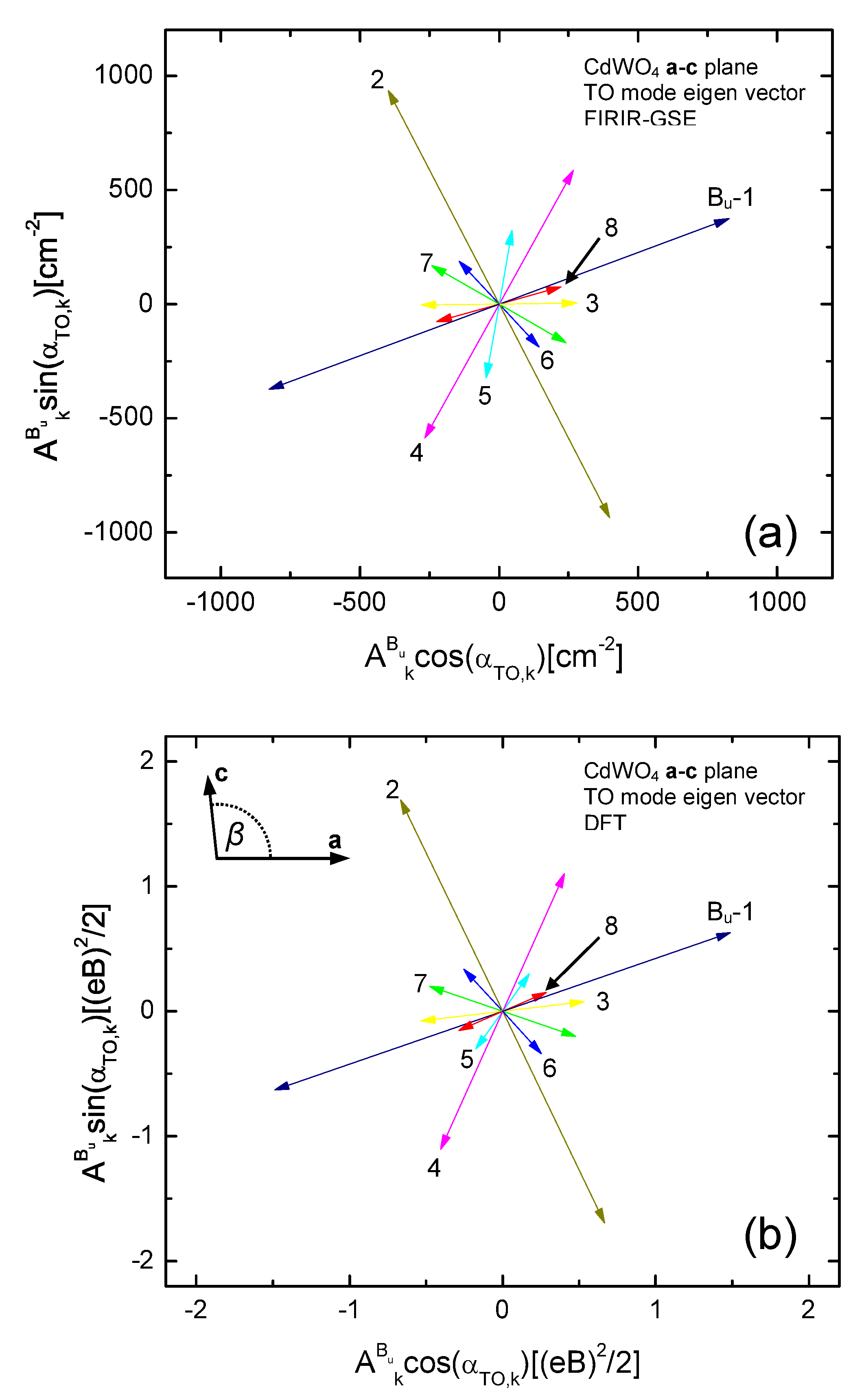}

    \caption{(a): Schematic presentation of the $B_u$ symmetry TO  mode eigen dielectric displacement unit vectors within the $\mathbf{a}$ - $\mathbf{c}$ plane according to TO mode amplitude parameters $A^{B_u}_k$ and orientation angles $\alpha_{\stext{TO},k}$ with respect to axis $\mathbf{a}$ obtained from GSE analysis (Tab.~\ref{tab:TOAuBuGSE}). (b) DFT calculated  $B_u$ mode TO phonon mode long wavelength transition dipoles (intensities) in coordinates of axes $\mathbf{a}$ and $\mathbf{c^{\star}}$ (Fig.~\ref{fig:CdWO4cell}).}
    \label{fig:acvec}
  \end{center}
\end{figure}
\paragraph{TO displacement unit vectors:} A schematic presentation of the oscillator function amplitude parameters $A^{B_u}_k$ and the mode vibration orientations according to angles $\alpha_{\stext{TO},k}$ from Tab.~\ref{tab:TOAuBuGSE} within the $\mathbf{a}$ - $\mathbf{c}$ plane is shown in Fig.~\ref{fig:acvec}(a). In
Fig.~\ref{fig:acvec}(b) we depict the projections of the DFT calculated long wavelength transition dipole moments (intensities) onto axes $\mathbf{a}$ and $\mathbf{c^{\star}}$, for comparison. Overall, the agreement is remarkably good between the TO mode eigen displacement vector distribution within the $\mathbf{a}$ - $\mathbf{c}$ plane obtained from GSE and DFT results. We note that the angular sequence of the $B_u$ mode eigen vectors follows those obtained by GSE analysis. Overall, the DFT calculated angles $\alpha$ agree to within less than 22$^{\circ}$ of those found from our GSE model analysis. Note that the eigen displacement vectors describe a uni-polar property without a directional assignment. Hence, $\alpha$ and $\alpha\pm\pi$ render equivalent eigen displacement orientations.

\begin{figure*}[!tbp]
  \begin{center}
      \includegraphics[width=0.99\linewidth,natwidth=4327,natheight=1675]{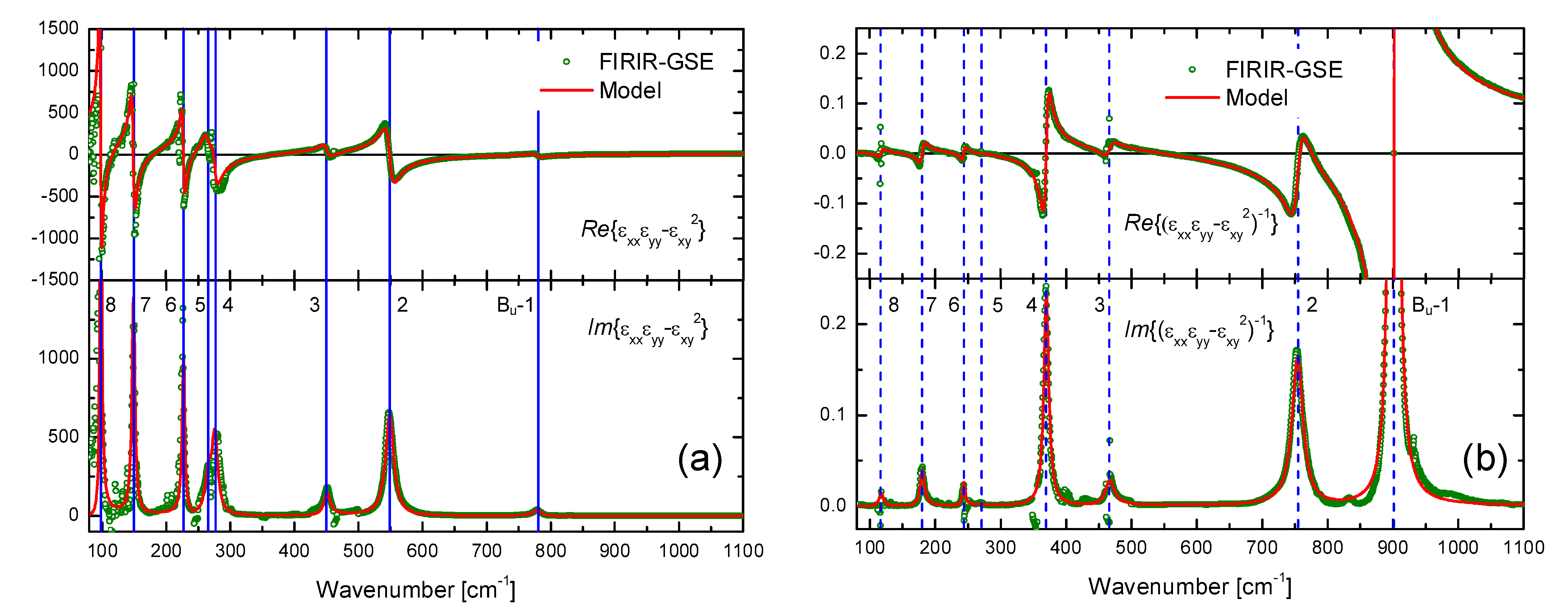}

    \caption{Real and imaginary parts of the coordinate invariant generalized monoclinic dielectric function $\varepsilon_{xx}\varepsilon_{yy}-\varepsilon_{xy}^{2}$ (left panels) and -($\varepsilon_{xx}\varepsilon_{yy}-\varepsilon_{xy}^{2}$)$^{-1}$ (right panels). Best-match model calculated data using the Berremann-Unterwald-Lowndes (BUL) form (solid lines) provide excellent match to ``experimental'' data (dotted lines) obtained from wavelength by wavelength generalized spectroscopic ellipsometry data analysis. Both TO and LO mode frequencies and broadening parameters can be determined, regardless of their unit vector orientation and amplitude parameters. Vertical lines indicate $B_u$ mode TO (dashed lines) and LO frequencies (dash dotted lines). Note that the imaginary parts of $\varepsilon_{xx}\varepsilon_{yy}-\varepsilon_{xy}^{2}$ and -($\varepsilon_{xx}\varepsilon_{yy}-\varepsilon_{xy}^{2}$)$^{-1}$ are found positive throughout the spectral range investigated.}
    \label{fig:detaceps}
  \end{center}
\end{figure*}

\paragraph{LO modes:} We use the generalized coordinate-invariant form of the dielectric function in Eq.~\ref{eq:general-eps-broaded} and match the function $\varepsilon_{xx}\varepsilon_{yy}-\varepsilon^{2}_{xy}$ obtained from the wavelength by wavelength obtained tensor spectra. All $B_u$ TO mode parameters, and parameters $\varepsilon_{\infty,xx}\varepsilon_{\infty,yy}-\varepsilon^2_{\infty,xy}$ are used from the previous step. Fig.~\ref{fig:detaceps} presents the imaginary parts of the functions $\varepsilon_{xx}\varepsilon_{yy}-\varepsilon_{xy}^{2}$, and $-(\varepsilon_{xx}\varepsilon_{yy}-\varepsilon_{xy}^{2})^{-1}$. The best-match model calculated data are obtained using the BUL form\cite{Berreman68,Lowndes70} to represent the coordinate invariant generalization of the dielectric function for materials with monoclinic symmetry, suggested in this present work. The presentation of the imaginary parts of the function and its inverse highlights the TO modes and LO modes as the broadened poles, respectively. The form results in an excellent match to the function calculated from the wavelength by wavelength experimental data analysis. Both TO and LO mode frequencies and broadening parameters can be determined, in principle, and regardless of their unit vector orientation and amplitude parameters. However, in our analysis here, we assumed values for all TO modes and only varied LO mode parameters, indicated by vertical lines in Fig.~\ref{fig:detaceps}. As a result, we find 8 LO modes with $B_u$ symmetry, and their broadening parameters, which are summarized in Tab.~\ref{tab:TOAuBuGSE}. An observation made in this work is noted by the spectral behavior of the imaginary parts of $\varepsilon_{xx}\varepsilon_{yy}-\varepsilon_{xy}^{2}$ and -($\varepsilon_{xx}\varepsilon_{yy}-\varepsilon_{xy}^{2}$)$^{-1}$, which are found always positive throughout the spectral range investigated. This suggests that the generalized coordinate-invariant form of the dielectric function in Eq.~\ref{eq:general-eps-broaded} (and the negative of its inverse) possesses positive imaginary parts as a result of energy conservation. A direct prove for this statement is not available at this point and will be presented in a future work.

The BUL form is used for analysis of functions $\varepsilon_{zz}$ and $\varepsilon^{-1}_{zz}$ for LO modes with $A_u$ symmetry. All TO mode parameters, and $\varepsilon_{\infty,zz}$ are used from the previous step. We find 7 LO modes, and their parameter values are summarized in Tab.~\ref{tab:TOAuBuGSE}. The best match calculated data and the wavelength-by-wavelength obtained spectra are depicted in Fig.~\ref{fig:epsilonmatrix} for $\varepsilon_{zz}$ (panel [d]), and Fig.~\ref{fig:epsiloninversematrix} for $\varepsilon^{-1}_{zz}$ (panel [d]).

\paragraph{Schubert-Tiwald-Herzinger condition:} The condition for the TO and LO broadening parameters in materials with multiple phonon modes and orthorhombic and higher crystal symmetry (Eq.~\ref{eq:STHcondition}) is fulfilled for polarization along axis $\mathbf{b}$ (See Tab.~\ref{tab:TOAuBuGSE}). The application of this rule for the TO and LO mode broadening parameters for phonon modes with their unit vectors within the monoclinic plane, and with general orientations in triclinic materials has not been derived yet. Hence, its applicability to modes with $B_u$ symmetry is speculative. However, we do find this rule fulfilled when summing over all differences between LO and TO mode broadening parameters in Tab.~\ref{tab:TOAuBuGSE}.

\paragraph{``TO-LO rule''} In materials with multiple phonon modes, a so-called TO-LO rule is commonly observed. According to this rule, a given TO mode is always followed first by one LO mode with increasing frequency (wavenumber). This rule can be derived from the eigen dielectric displacement summation approach when the unit vectors and functions $\varrho_{l}$ possess highly symmetric properties. A requirement for the TO-LO rule to be fulfilled can be suggested here, where the TO and LO modes must possess parallel unit eigendisplacement vectors. For example, this is the case for polarization along axis $\mathbf{b}$, hence, the TO-LO rule is found fullfilled for the 7 pairs of TO and LO modes with $A_u$ symmetry. For the TO and LO modes with $B_u$ symmetry, none of their unit vector is parallel to one another, hence, the TO-LO rule is not applicable. For monoclinic $\beta$-Ga$_2$O$_3$ we observed that the rule was broken. The explanation was given by the fact that the phonon mode eigendisplacement vectors are not parallel within the $\mathbf{a}-\mathbf{c}$ plane.\cite{Schubert_2016} Nonetheless, we note that the rule is not broken for CdWO$_4$. Whether or not the TO-LO rule is violated in a monoclinic (or triclinic) material may depend on the strength of the individual phonon mode displacement amplitude and their orientation.

\begin{table}[!t]
\caption{Best match model parameters for high frequency dielectric constants. The static dielectric constants are obtained from extrapolation to $\omega=0$. The S-LST relation is found valid with TO and LO modes given in Tab.~\ref{tab:TOAuBuGSE}.}
\begin{center}
\begin{tabular}{lcccc}
\noalign{\bigskip} \hline \hline
\noalign{\smallskip}
& $\varepsilon_{xx}$ ($\mathbf{a}$) & $\varepsilon_{yy}$ ($\mathbf{c^{\star}}$)& $\varepsilon_{yx}$ & $\varepsilon_{zz}$($\mathbf{b}$)\\
\noalign{\smallskip}
 \hline
 \noalign{\smallskip}
$\varepsilon_{\infty,(j)}$& 4.46(1)&4.81(1)&0.086(6)&4.25(1) \\
$\varepsilon_{\stext{DC},(j)}$& 16.16(1)&16.01(1)&1.05(1)&11.56(1)\\
\hline
\end{tabular}
\end{center}\label{tab:fccepsDCinf}
\end{table}

\paragraph{Static and high frequency dielectric constant:} Tab.~\ref{tab:fccepsDCinf} summarizes static and high frequency dielectric constants obtained in this work. Parameter values for $\varepsilon_{\stext{DC}}$ were estimated from extrapolation of the tensor elements in the wavelength-by-wavelength determined $\varepsilon$. Values for $\varepsilon_{\stext{DC,xx}}$ and $\varepsilon_{\stext{DC,yy}}$ agree well with the value of 17 given by Shevchuk and Kayun\cite{Shevchuk_2007} measured at 1~kHz on a (010) surface. We find that with the data reported in Tab.~\ref{tab:TOAuBuGSE} and Tab.~\ref{tab:fccepsDCinf} the S-LST relation in Eq.~\ref{eq:SLST} is fulfilled.

\section{Conclusions}

A dielectric function tensor model approach suitable for calculating the optical response of monoclinic and triclinic symmetry materials with multiple uncoupled long wavelength active modes was applied to monoclinic CdWO$_4$ single crystal samples. Surfaces cut under different angles from a bulk crystal, (010) and (001), are investigated by generalized spectroscopic ellipsometry within mid-infrared and far-infrared spectral regions. We determined the frequency dependence of 4 independent CdWO$_4$ Cartesian dielectric function tensor elements by matching large sets of experimental data using a polyfit, wavelength-by-wavelength data inversion approach. From matching our monoclinic model to the obtained 4 dielectric function tensor components, we determined 7 pairs of transverse and longitudinal optic phonon modes with $A_u$ symmetry, and 8 pairs with $B_u$ symmetry, and their eigenvectors within the monoclinic lattice. We report on density functional theory calculations on the mid-infrared and far-infrared optical phonon modes, which  are in excellent agreement with our experimental findings. We also discussed and presented monoclinic dielectric constants for static electric fields and frequencies above the reststrahlen range, and we observed that the generalized Lyddane-Sachs-Teller relation is fulfilled excellently for CdWO$_4$.

\section{Acknowledgments} This work was supported in part by the National Science Foundation (NSF) through the Center for Nanohybrid Functional Materials (EPS-1004094), the Nebraska Materials Research Science and Engineering Center (MRSEC) (DMR-1420645) and awards CMMI 1337856 and EAR 1521428. The authors further acknowledge financial support by the University of Nebraska-Lincoln, the J.~A.~Woollam Co., Inc., and the J.~A.~Woollam Foundation. Parts of the DFT calculations were performed using the resources of the Holland Computing Center at the University of Nebraska-Lincoln.

\bibliography{CompleteLibrary_RK}

\begin{thebibliography}{72}
\bibitem [{foo({\natexlab{b}})}]{footnotePRBalyssa2}%
  \BibitemOpen
  \href@noop {} {\bibfield  {journal} {\bibinfo  {journal} {One may simply
  observe the spectral behavior of the imaginary parts only, and identify from
  there the locations of absorption ``bands'' and identify their center
  frequencies.}\ } ({\natexlab{b}})}\BibitemShut {NoStop}%
\end{thebibliography}
\end{document}